\documentclass[10pt,prl,preprint,a4paper,twocolumn,showpacs,superscriptaddress]{revtex4}
\usepackage{graphicx}

\newcommand{\f}{\frac}
\newcommand{\be}{\begin{equation}}
\newcommand{\ee}{\end{equation}}
\newcommand{\bea}{\begin{eqnarray}}
\newcommand{\eea}{\end{eqnarray}}

\newcommand{\Jeff}{J_{\rm eff}}
\newcommand{\down}{-}
\newcommand{\up}{+}

\begin{document}

\title{Anomalous Spin Excitation Spectrum of the Heisenberg Model in a Magnetic Field}

\author{Olav F. Sylju{\aa}sen}
\email{sylju@nordita.dk}
\affiliation{NORDITA, Blegdamsvej 17, DK-2100 Copenhagen {\O},
         Denmark}

\author{Patrick A. Lee}
\affiliation{Department of Physics, Massachusetts Institute of Technology, Cambridge, MA 02139}
%\thanks{Thanks to....}

\date{\today}


\preprint{NORDITA-2001-29CM}

\begin{abstract}
Making the assumption that high energy fermions exist in the two dimensional 
spin-1/2 Heisenberg antiferromagnet we present predictions based on the $\pi$-flux ansatz for the dynamic structure factor when the antiferromagnet is subject to a uniform magnetic field. The main result is the presence of gapped excitations in a momentum region near $(\pi,\pi)$ with energy lower than that at $(\pi,\pi)$. This is qualitatively different from spin wave theory predictions and may be tested by experiments or by quantum Monte Carlo.
\end{abstract}

\maketitle

It is well known in particle physics 
that the nature of elementary excitations for one and the same model
might change as different
momenta- and energy- scales are probed. 
In particular, mesons at very high energies are best described
as pairs of fermions (quarks) while at low energies they are bosons. 
It has been suggested that a similar possibility exists in the Heisenberg 
spin-1/2 antiferromagnet on a square lattice,
where the low energy excitations are Goldstone bosons while the high energy excitations show features more resembling fermions (loosely called spinons in the literature). 
Taking this as a real possibility we describe here the consequences of adding a magnetic
field to the model in order to bring out clearly the signatures of the underlying high energy fermions. The predictions made in this paper can be tested experimentally in materials such as copper formate tetra-deuterate (CFTD) or the organic compound (5CAP)2CuBr4 where the exchange constant is not too large. They can also be tested using numerical tools such as quantum Monte Carlo.

The recently observed dispersion along the antiferromagnetic zone boundary
$(\pi/2,\pi/2)$-$(0,\pi)$ (lattice spacing is set to unity throughout this Letter) in experiments on CFTD\cite{Ronnow} is an indication that the nature of high-energy excitations in the Heisenberg model might be different from the low energy ones. 
This dispersion, which also has
been confirmed using Monte Carlo simulations\cite{Me}, 
shows a broad shallow minimum around $(0,\pi)$. 
This feature which is not captured within linear spin wave theory (LSW) is reminiscent of a prediction made by Hsu\cite{Hsu}. Treating the Heisenberg model in an approximation involving massive {\em fermions}, he found a rather deep minimum in the dispersion at $(0,\pi)$. While this minimum seems much to deep to explain the experiments,
Ho {\em et al}.\cite{Ho} recently calculated the {\em full} dynamic structure factor using the same fermionic picture. They showed that the pole at $(0,\pi)$ merges with an extensive high-energy continuum carrying lots of spectral weight. For experiments this implies that the spectral peak at $(0,\pi)$ would be very broad and an accurate determination of the magnon energy would be difficult, if anything, the high-energy continuum would make the dispersion appear shallower than in Hsu's original prediction. 

Neutron scattering is necessarily an indirect probe of spin-1/2 fermions as it measures spin-1 excitations. So in order to establish the existence of fermionic excitations it appears that one need to be able to distinguish between sharp peaks(magnons) and broad continua(convolution of fermions). However, as will be shown here one can bring out the difference more clearly by applying a uniform magnetic field.
This is known in the context of the 1d Heisenberg antiferromagnet where the elementary excitations are true spinons.  There the dynamic structure factor in a magnetic field shows low energy peaks at incommensurate momentum-values which change with the magnitude of the applied magnetic field. This effect which has been seen experimentally\cite{Dender} follows from the Bethe ansatz solution, but can also be understood as a splitting of the mean-field spin up- and down- spinon-bands\cite{Pytte}. This splitting causes new nodes to appear at values determined by the magnetic field thereby causing low-energy peaks in the dynamic structure factor resulting from inter-node scattering. 
 
The formulation of the 2d Heisenberg model in terms of fermions is based on the $\pi$-flux ansatz\cite{Affleck} where first each spin-operator is written as a pair of fermions obeying the constraint of one particle per lattice site. Then the resulting quartic fermion coupling is decoupled with a mean field hopping term with imaginary parts such that the phase change in moving a fermion around an elementary plaquette is $\pi$. This mean field theory must be supplemented by gauge fields
which enforce the single occupancy constraint. An alternative approach is
to use the mean field theory as the starting point for constucting a
variational wavefunction. In the Gutzwiller projected wavefunction, the
doubly occupied and unoccupied states are removed by hand, thereby
satisfying the constraint exactly. While the Gutzwiller projection of the
$\pi$-flux state gives quite good energy, the best trial wavefunction is one
which combines the $\pi$-flux state and a spin density wave state. The
projection of this state gives long range Neel order with a sublattice
magnetization which is in excellent ageement with exact computations.\cite{Hsu}
Thus we are led to consider an effective Hamiltonian whose mean field
solution gives the $\pi$-flux state with the spin density wave. This is
accomplished by introducing a weak on-site repulsion V. The value of V is
to be determined variationally to produce the proper magnetization and
should not be confused with the original strong Hubbard U. Adding also a 
term coupling the fermions to the magnetic field the resulting effective Hamiltonian reads
\bea
	{\cal H} &=& \Jeff \left\{ -\f{1}{2} \sum_{<ij>} \left( e^{i \phi_{ij}} f^\dagger_{i \sigma} f_{j \sigma} + h.c. \right) 
	 	\right. \nonumber \\
	&+&   
	\f{V}{\Jeff} \sum_i  \left( n_{i \up}-\f{1}{2} \right) \left( n_{i \down}-\f{1}{2} \right) \label{Heff} \\
 	&+&
	\left.
	\f{h}{2} \sum_i \left( f^\dagger_{i \up} f_{i \down} 
	                       + \rm{ h.c.} \right) 
	\right\}, \nonumber
\eea 
where $f^\dagger_{i\sigma}$ denotes creation of a fermion at site $i$ with spin projection $\sigma=\{-,+\}$ along the $S^z$-axis, $n_{i\sigma}$ is the corresponding density operator and $\phi_{ij}$ is a phase on each link on the lattice chosen such that when summed clockwise around an elementary plaquette the sum equals $\pi \; {\rm mod} (2\pi)$.   
As the fermions carry no charge we have coupled the magnetic field as a Zeeman term to $S^x$. We must choose $S^x$ (or $S^y$, not $S^z$), as coupling to $S^z$ cannot be realized experimentally, for even in a very small field the antiferromagnet will orient itself such that the staggered direction, which here is chosen to be the $z$-direction, is perpendicular to the applied field. 

\begin{figure}[!t]
\includegraphics[clip,width=0.7\columnwidth]{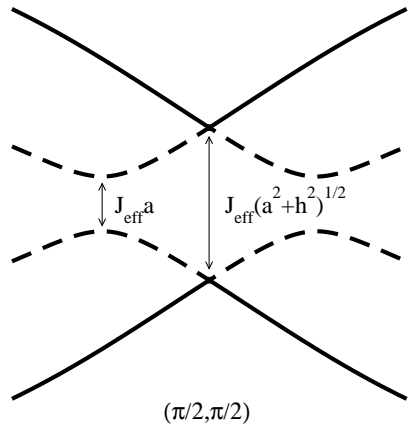}
\caption{Fermion band dispersion in the presence of a magnetic field along the zone diagonal around the point $(\pi/2,\pi/2)$. Dashed line indicates branch $1$ while branch $2$ is solid. \label{bands}}
\end{figure}

The repulsive potential will be treated self-consistently. That is, the potential term is first decoupled writing
\be
	<n_{i\sigma}> = \f{1}{2} + \sigma (-1)^i m, \label{SCE1}
\ee
where $m$ is the staggered magnetization. The resulting quadratic Hamiltonian is then diagonalized and used to determine the left hand side of Eq.~(\ref{SCE1}), which then gives a relation between $V/\Jeff$, $m$ and $h$,
\be
	V^{-1} = \f{2}{N} \sum^\prime \f{1}{4} \left( \f{1}{E_1} + \f{1}{E_2} \right), \label{selfcons}
\ee
where the sum is taken over the antiferromagnetic Brillouin zone, $N$ is the total number of sites and $E_1$,$E_2$ are the dispersions gotten by diagonalizing the quadratic Hamiltonian,
\be
	E_{\stackrel{1}{\scriptscriptstyle 2}} = \Jeff \sqrt{ (\epsilon_k \mp h/2)^2 + (a/2)^2 },
\ee 
where $\epsilon_k = (\cos^2 k_x+\cos^2 k_y)^{1/2}$, and $a=mV/\Jeff$, see Fig.~\ref{bands}.

To make the connection to the parameters in the original Heisenberg model: the exchange constant $J$ and the magnetic field $H$, 
we will consider the Gutzwiller projected ground state of the quadratic Hamiltonian above as a variational ansatz for the Heisenberg model ground state. 
In calculating the Gutzwiller projection we proceed as Hsu\cite{Hsu} and calculate the projection exactly on a small cluster while approximating longer range correlations using combinatorics. While
Hsu used a $2\times2$ cluster the results quoted here is obtained from a 
$3\times 3$ cluster.

\begin{center}
\begin{figure}[ht]
\includegraphics[clip,width=0.9\columnwidth]{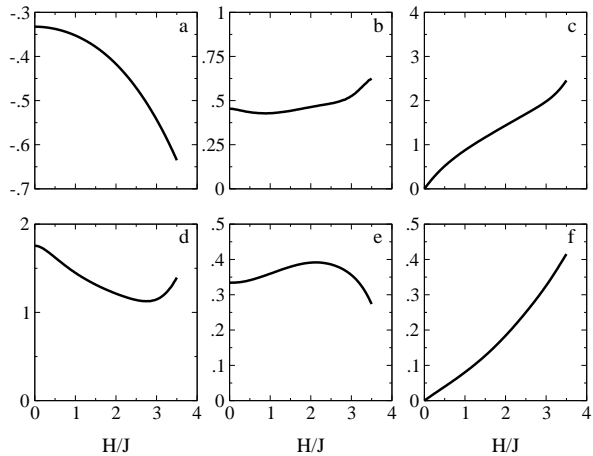}
\caption{Parameters determined from the Gutzwiller projection on a $3 \times 3$ cluster. The graphs show: (a) energy, (b) $a$, (c) $h$, (d) $V/\Jeff$, (e) Staggered magnetization and (f) Total magnetization as functions of the applied magnetic field\label{Gutzwiller} }
\end{figure}
\end{center}

Expressing the spin operators as fermions, which again can be expressed as linear combinations of the eigenstates of the quadratic Hamiltonian, the Gutzwiller projected energy of the Heisenberg model in a fixed magnetic field H is minimized with respect to the parameters $h$ and $a$. Having obtained these values $V/\Jeff$ is calculated from Eq.~(\ref{selfcons}). This procedure does not fix the overall scale $\Jeff$ which indeed was taken to be the bare value $J$  in Hsu's original work. Instead we will adjust $\Jeff$ to fit the spin wave velocity as discussed below.
In Fig.~\ref{Gutzwiller} we show the values of $a$,$h$ and $V/\Jeff$ as functions of the magnetic field obtained in this way. We have also plotted the total energy and the staggered and uniform magnetization after projection.
It should be emphasized that this procedure of finding the effective
parameters contains no adjustables.
In zero magnetic field the obtained value of $a$ is 0.45 which is close to the value 0.5 used in Ref.\cite{Ho}.
Furthermore the energy and staggered magnetization are $-0.333J$ and 
$0.33$ respectively which should be compared to $-0.334671J$ and 
$0.3070(3)$ obtained from quantum Monte Carlo simulations\cite{Sandvik}. It should be mentioned that the values gotten from this Gutzwiller-projected method is approximate in the sense that a finite-sized cluster is used and no extrapolation to the infinite system have been attempted.  
From Fig.~\ref{Gutzwiller} it is seen that for small magnetic fields $a$ stays roughly constant while $h$ increases approximately linearly with $H/J$. 
For extremely high magnetic fields close to the saturation field $H/J=4$, the minimization procedure cannot alone determine values of $a$ and $h$ as there are many values of $a$ and $h$ maximizing the magnetization, which is the overwhelming contribution to the energy at high fields. We have therefore omitted the narrow range from $H/J=3.5$ and up to the saturation field.

To calculate the dynamic structure factor we calculate the fermion pair susceptibility and employ the RPA approximation using the values of the parameters gotten in Fig.~\ref{Gutzwiller}. 
As shown in Ref.~\cite{Ho} this gives the correct Goldstone behavior and also gives information about the full spectrum. 
At zero temperature the dynamic structure factor is $S(\omega)=-2{\rm Im}\chi(\omega_n=\omega+i\delta)$ where using the RPA we have
\be
	\chi(\omega_n) = \f{\chi_o(\omega_n)}{1-{\cal V} \chi_o(\omega_n)}, \label{RPA}
\ee
where all the quantities are matrices with spin and momentum space indices. $\chi_o(\omega_n)$ is a short-hand for $\chi_{o,\alpha \beta ;\gamma \delta}(q,q^\prime,\omega_n)$ which is a convolution of the fermion propagators $G_{\beta \gamma}$ and $G_{\delta \alpha}$,
where $G_{\alpha \beta}(k,k^\prime,\omega_n)=-<f_{\beta,k^\prime,\omega_n} f^\dagger_{\alpha,k,\omega_n}>$, and  $\alpha$,$\beta$ are spin indices. ${\cal V}$ is diagonal in momentum-space and the non-zero elements in spin space are ${\cal V}_{\up \up ;\down \down}={\cal V}_{\down \down ; \up \up}=-{\cal V}_{\up \down; \down \up}=-{\cal V}_{\down \up; \up \down}=V$. In the presence of a staggered magnetization it is important to note that $\chi_o$ has non-zero off-diagonal Umklapp elements which will contribute to the momentum-space diagonal elements of $\chi$. The calculation follows closely that in Ref.~\cite{Schrieffer}, and the detailed results will be presented elsewhere. Here we will give the qualitative details as well as the most important numerical results. 

\begin{center}
\begin{figure}[!ht]
\includegraphics[width=\columnwidth]{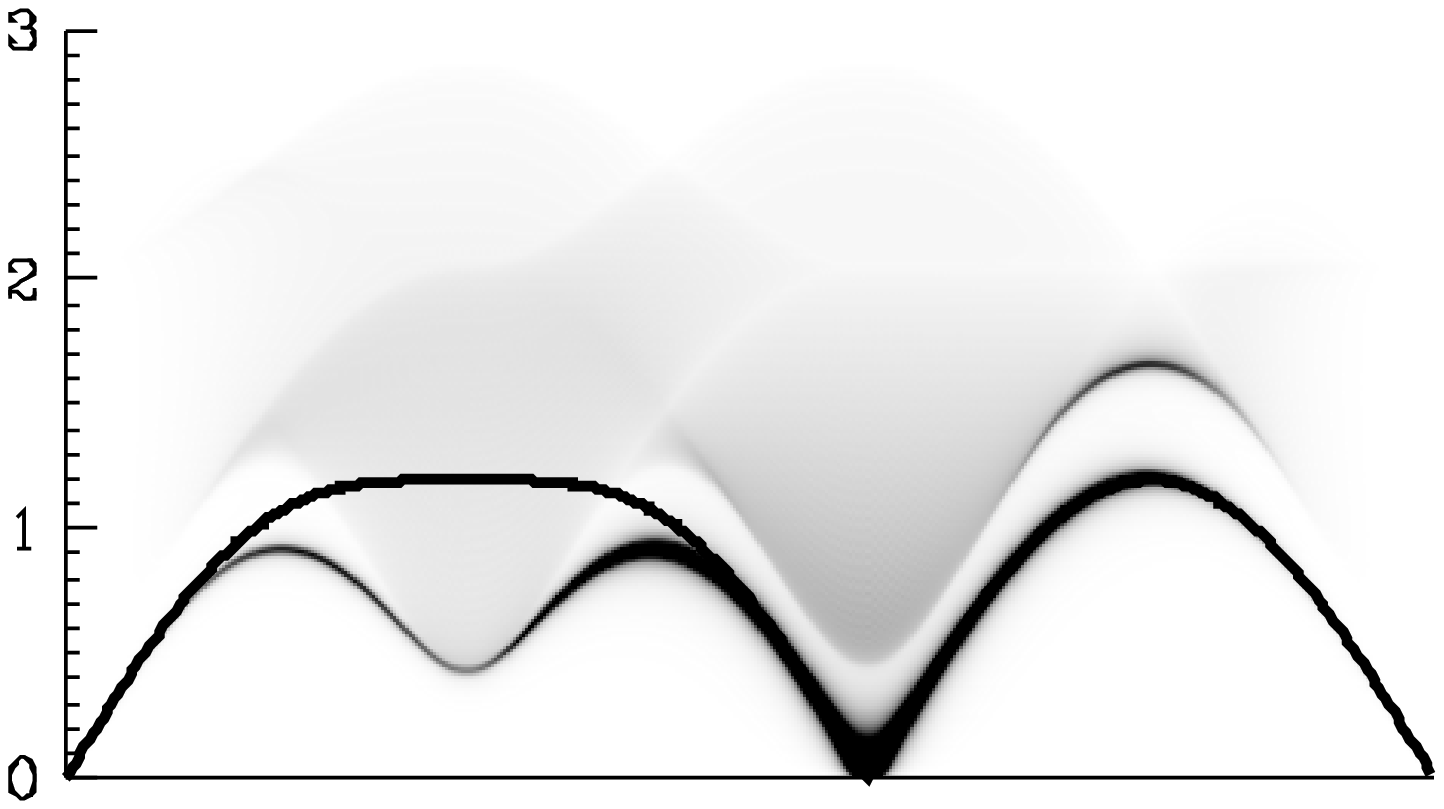}
\includegraphics[width=\columnwidth]{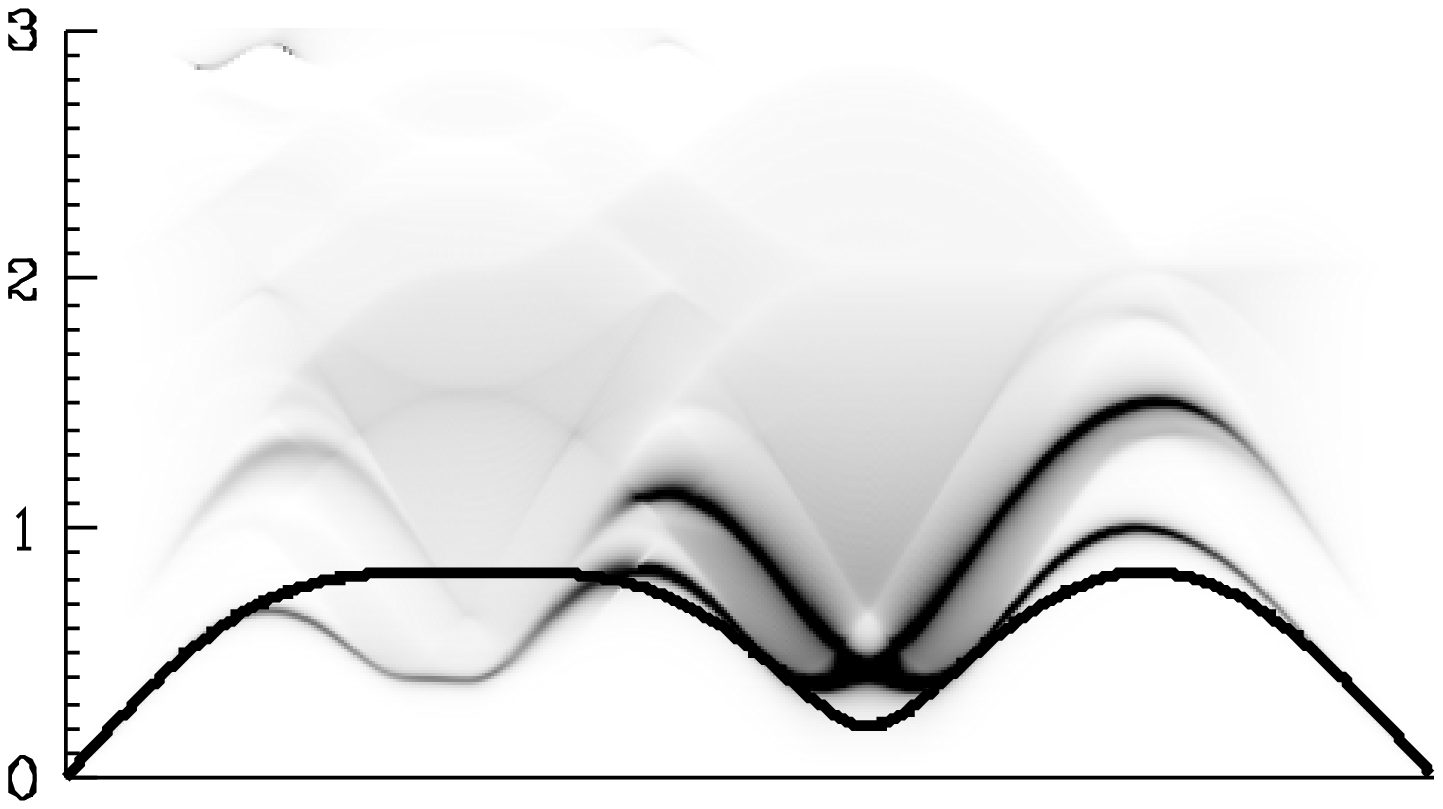}
\includegraphics[width=\columnwidth]{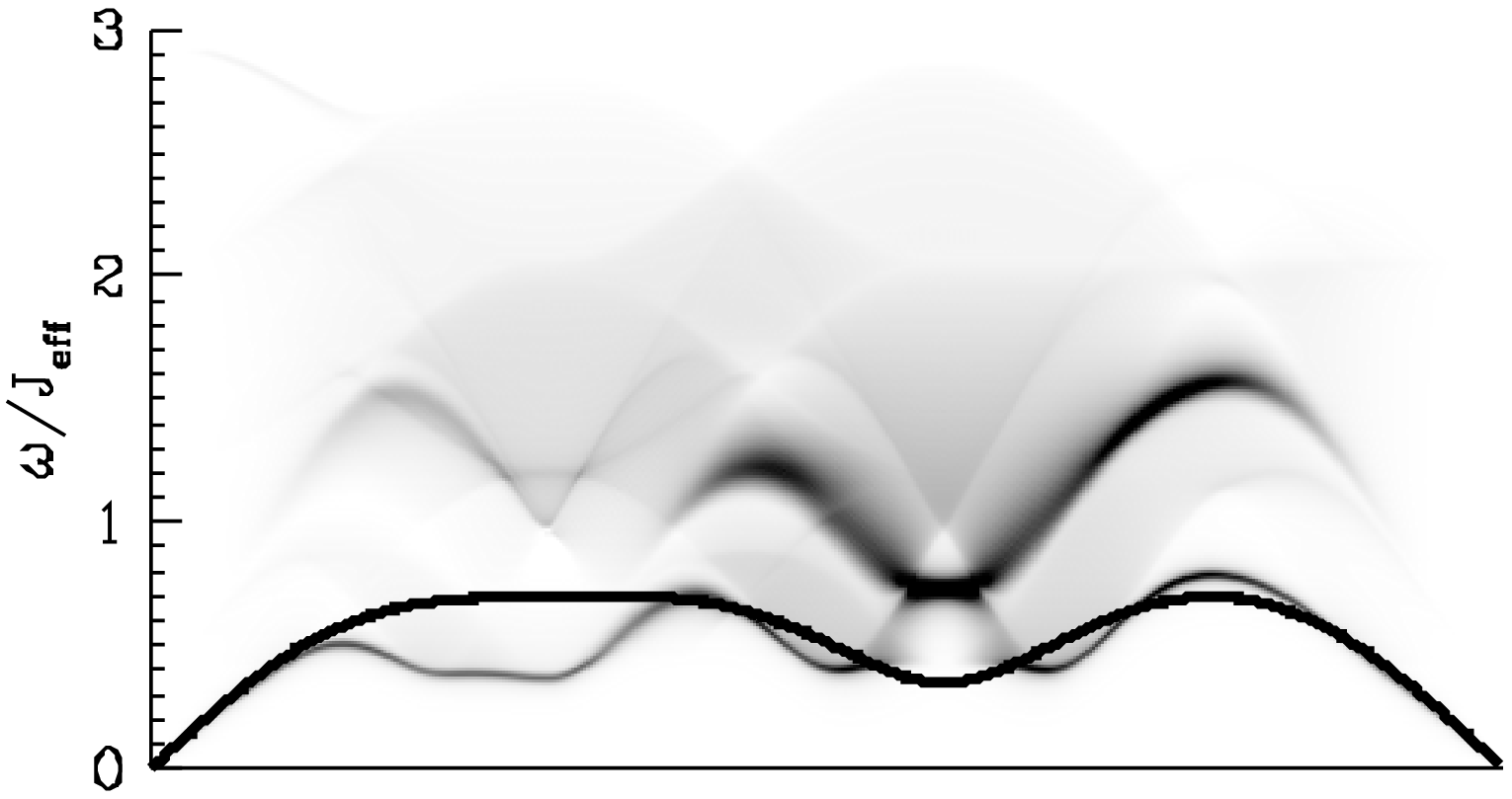}
\includegraphics[width=\columnwidth]{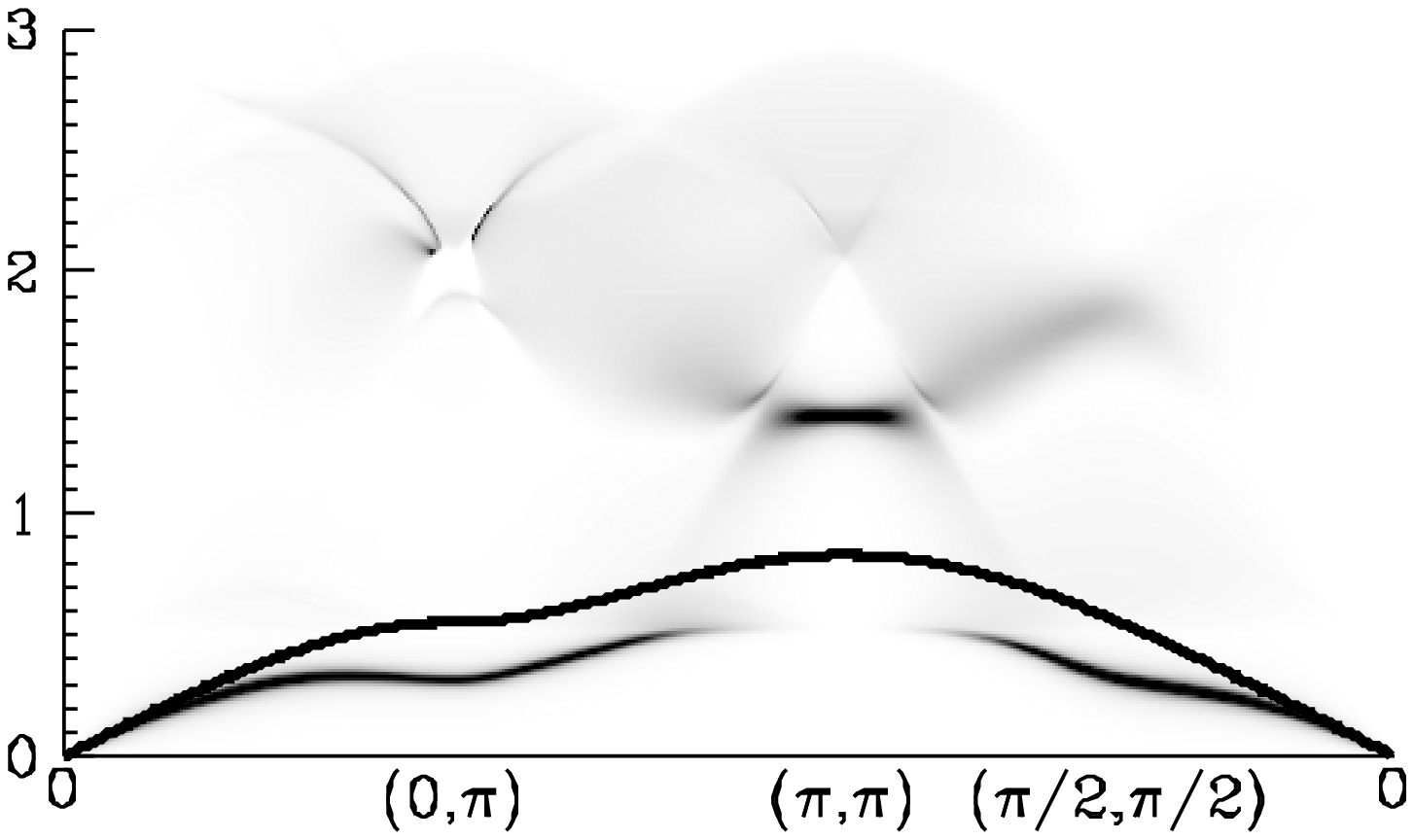}
\caption{Gray-level plots of the dynamic structure factor $S^{xx}$ along the zone edges and diagonal for parameters corresponding to, from top to bottom, $H/J = 0.0,0.5,1.0$ and $3.0$. The solid line is the LSW result, where we have used, from top to bottom, $\Jeff/J= 1.67$, $2.43$, $2.87$ and $3.61$.\label{rparesults}}
\end{figure}
\end{center}
Numerical results for the dynamic structure factor $S^{xx}$, that is for spin components {\em along} the magnetic field, for four different values of the magnetic field are shown in Fig.~\ref{rparesults}.
The top panel which is for the zero field case shows clearly the deep minimum around $(0,\pi)$ discovered by Hsu\cite{Hsu} as well as the merging of the high energy continuum with the dominant spectral peak at this point calculated in Ref.~\cite{Ho}.\cite{spin-wave-vel} The thick solid line is the spin wave spectrum according to LSW\cite{Zfactor}.
From the second panel in Fig.~\ref{rparesults} we see that for small fields the Goldstone mode at $(\pi,\pi)$ becomes gapped\cite{nogap} in accordance with LSW. However, unlike LSW, the spectral function splits into a double peak structure near $(\pi,\pi)$, and the energy in an incommensurate ring around $(\pi,\pi)$ lies slightly below that at $(\pi,\pi)$. On further increasing the magnetic field this energy difference between the incommensurate ring and the peak at $(\pi,\pi)$ increases as can be seen in the third panel in Fig.~\ref{rparesults} where $H/J=1$.

The incommensurate low-energy peaks at finite fields come from the 1-1-fermion scattering which 
also is responsible for the remnants of the zero field minimum at $(0,\pi)$.
For a finite magnetic field and $h < 2\sqrt{2}$, which is the case almost up to saturation branch 1 develops characteristic energy extrema, as indicated in Fig.~\ref{bands}, with energies $\pm \Jeff a/2$ along the contours $( \sin^2(r_x) + \sin^2(r_y) )^{1/2} =h/2$, where $\vec{r}$ is the displacement from the $(\pi/2,\pi/2)$ and
symmetric points, while branch $2$ keeps its extremal values $\Jeff (a^2+h^2)^{1/2}/2$ 
at $(\pi/2,\pi/2)$ and symmetric points. The fact that the energy extremal values for branch 1 are independent of $h$ for $h < 2\sqrt{2}$ has important consequences as it implies the presence of relatively low-energy excitations $\sim \Jeff a$ due to 1-1 fermion scattering over almost the whole magnetic field range. While the energies of these excitations do not depend on $h$, they depend on $H$ thru $\Jeff$ ($a$ is only weakly dependent on $H$).

To explain why the low energy structure around $(\pi,\pi)$ is 
lower than at $(\pi,\pi)$ one can consider the
coherence factors which appears in the calculation of 
$\chi_o^{xx}$. These coherence factors $C_{11}$, $C_{12}$ and $C_{22}$, associated with the three different types of scattering, are proportional to $(1+g)$, $(1-g)$ and $(1+g)$ respectively, 
%\bea
%	C_{11} & = & (1+g)\left( 1-
%	               \f{(\epsilon_k-h/2)(\epsilon_{k+q}-h/2)-(a/2)^2}
%			{E_{1}(k) E_{1}(k+q)} \right) \nonumber \\
%	C_{12} & = & (1-g)\left( 1+
%	               \f{(\epsilon_k-h/2)(\epsilon_{k+q}+h/2)+(a/2)^2}
%			{E_{1}(k) E_{2}(k+q)} \right) \nonumber \\
%	C_{22} & = & (1+g)\left( 1-
%	               \f{(\epsilon_k+h/2)(\epsilon_{k+q}-h/2)-(a/2)^2}
%			{E_{2}(k) E_{2}(k+q)} \right) \nonumber
%\eea
where 
\be
 g = \f{\cos(k_x) \cos(k_x+q_x)+\cos(k_y) \cos(k_y+q_y)}{\epsilon_k \epsilon_{k+q}}.
	\nonumber
\ee
For $q$, the momentum of the pair-excitation equal 0, $g=1$, while it is $-1$ for $q=(\pi,\pi)$. Because of this, 1-1-fermion scattering will be suppressed in the region very close to $(\pi,\pi)$ and the dominant spectral peak at $(\pi,\pi)$ will follow the characteristic energy of 1-2-fermion scattering which
depends on $h$. Away from $(\pi,\pi)$, 1-1-scattering, which does not depend on $h$, reappears giving rise to the low energy ring-like structure.   

We have plotted the LSW result for comparison, where we have determined $\Jeff$ so that the velocity of the remaining Goldstone boson equals the spin-wave velocity in LSW. As can be seen from Fig.~\ref{rparesults} LSW with $\Jeff$ determined this way gives consistently a lower value for the gap at $(\pi,\pi)$ than is gotten using the fermion picture\cite{Zhitomirsky}. However, independently of the value of $\Jeff$, the fermion-result is still very different from LSW, especially the presence of low energy excitations (lower than at ($\pi,\pi$)) in an incommensurate ring around $(\pi,\pi)$.   

At high magnetic fields the phase space for 1-1- and 1-2-scattering are well separated in energy, and so for high fields the energy position of the dominant spectral peak changes very rapidly close to $(\pi,\pi)$. This is clearly seen for $H/J=3$ in the bottom panel of Fig. \ref{rparesults}.   

We hope that the predictions presented here can be checked in the near future,
either using neutron scattering experiments or numerical calculations. Observation of the anomalous features will provide strong support for the fermion formulation of the Heisenberg model.

\begin{acknowledgments}
P.A. Lee acknowledges the support from NSF grant DMR-9813764.
\end{acknowledgments}

%\end{multicols}

\begin{thebibliography}{00}
\bibitem{spin-wave-vel} Our results for the spin-wave velocity is 
$0.84\Jeff$. This value differs from the value $1.14 \Jeff$ gotten in \cite{Ho}. We attribute this to their choice of $a=0.5$ and their neglect of Umklapp processes which lower the spin-wave velocity.
\bibitem{Zfactor} $\Jeff$ is gotten by demanding that the spin-wave velocity extracted from plots of $S^{yy}$ close to $(\pi,\pi)$ matches the LSW result $c=J\sqrt{2-(H/J)^2/8}$. Any multiplicative uniform normalization $Z_c$ of 
the LSW spectrum coming from higher orders in $1/S$ will increase the value
of $\Jeff$ given here by a factor $Z_c$.
\end{thebibliography}
\end{document}